\documentstyle[preprint,aps]{revtex}

\def\gapproxeq{ $\,$ \raisebox{-.5ex}{$
\stackrel{>}{\scriptstyle\sim} $ } }
\def\lapproxeq{$\,$ \raisebox{-.5ex}{$
\stackrel{<}{\scriptstyle\sim } $ } }

\begin{document}

\title{Correlations in Nuclear Arrhenius-Type Plots}

\author{M. B. Tsang and P. Danielewicz}
\address{National Superconducting Cyclotron
Laboratory and Department of Physics and Astronomy, Michigan State
University, East Lansing, MI 48824, USA,}

\date{\today}
\maketitle

\begin{abstract}

Arrhenius-type plots for multifragmentation process, defined as the
transverse energy dependence of the single-fragment emission-probability, $
(\ln (1/p_{b})$ vs $ 1/\sqrt{E_{t}})$, have been studied by examining the
relationship of the parameters $p_{b}$ and $E_{t}$ to the intermediate-mass
fragment multiplicity $\langle n\rangle .$ The linearity of these plots reflects the
correlation of the fragment multiplicity with the transverse energy. These
plots may not provide thermal scaling information about fragment production
as previously suggested.

\end{abstract}
\pacs{PACS number: 25.70.Pq}
\newpage

About a hundred years ago, the Swedish chemist Svante Arrhenius discovered
that the rate of chemical reactions increases with temperature [1].
Specifically, the chemical reaction rate constant $(k)$ is related to the
absolute temperature $(T)$ by
\begin{equation}
k \propto \exp (-E_a/T)
\end{equation}
where $E_a$ is the activation energy of the chemical reaction. The linear
relationship between $\ln (k)$ and $1/T$ is widely known as Arrhenius plot
in chemistry and is associated with thermal equilibrium processes.

Recently, Arrhenius-type plots have been used to study the statistical [2-5]
and dynamical [6] properties of fragment emissions in heavy ion reactions.
In the intermediate incident energy range, between few tens of MeV to 100
MeV per nucleon, the production of intermediate mass fragments (IMF, 3 $\leq
$ $Z\leq 20)$, also known as multifragmentation, is an important decay mode
of highly excited nuclear systems [7]. Calculations suggest that the
fragments are produced in the phase co-existence region. Thus, understanding
the mechanisms of fragment formation may provide an insight to the liquid
gas phase transition in nuclear matter [7].

Several experiments [2-4] gave evidence that in collisions characterized by
a given value of the total transverse energy of detected charged particles,
\begin{equation}
E_{t}=\sum_{i}E_{i}\sin ^{2}\theta _{i},
\end{equation}
the IMF multiplicity distribution may be fitted by a binomial distribution,
\begin{equation}
P_{n}^{m}=\frac{m!}{n!(m-n)!}p_{b}^{n}(1-p_{b})^{m-n},
\end{equation}
where $n$ is the IMF multiplicity and the parameter $m$ is interpreted as
the number of times the system tries to emit a fragment. The probability of
emitting fragments can be reduced to a single particle emission probability $
p_b${\bf } for true binomial distributions; the binomial parameters $m$ and
$p_b$ are related to the mean and variance of the fragment multiplicity
distributions according to:
\begin{equation}
\langle n\rangle  = mp_b
\end{equation}
\begin{equation}
\sigma ^2_n=\langle n\rangle (1-p_b)
\end{equation}
The past investigations found a simple linear relationship between $\ln
(1/p_{b})$ and $1/$ $\sqrt{E_{t}}$ (nuclear Arrhenius-type plots), for
several projectile-target combinations and incident energies [2-4]. By
assuming a linear relationship between $\sqrt{E_{t}}$ and temperature $T$
and from the linearity of the observed $\ln (1/p_{b})$ vs $ 1/\sqrt{E_{t}}$
plot, it has been inferred that a thermal scaling of the multifragment processes
might be a general property [2-4]. In this picture, the observed ``linear"
 dependence of $\ln (1/p)$ upon $1/T$ would be reflecting the
\begin{equation}
p \propto \exp (-B/T)
\end{equation}
dependence of fragment emission probabilities upon a common fragment
emission barrier~$B$.

However, unlike chemical reactions, $p$ and $T$ were not measured directly in
Refs. [2-4]. There, the validity of the Arrhenius-type plots relies on two
assumptions: 1) that $E_{t}$ is proportional to the excitation energy $E^{*}
$, and therefore, should be proportional to $T^{2}$, and 2) that $p_{b}$
obtained by fitting the fragment multiplicity distributions is the
elementary emission probability $p$. This article will examine the above
assumptions and investigate the underlying reasons for the linearity
exhibited by the Arrhenius-type plots obtained in many systems.

The assumption that temperature is proportional to the square root of the
excitation energy, $T\propto \sqrt{E^{*}}$, is valid for compound nuclei
formed at low to moderate temperature. Some relationship between the
transverse energy and excitation energy and therefore temperature may also
be obtained for compound nuclei, provided adjustment is made for the Coulomb
barrier and for neutron emission [8]. For the intermediate-energy
heavy ion reactions such as $^{36}Ar+^{197}Au$ at $E/A=35$ to 110 MeV, where
linearity of the Arrhenius-type plots have been observed, however, the final
states contain fast particles emitted from the overlap region of projectile
and target as well as delayed emission from projectile- and target-like
residues. In particular, particle production from the overlap participant
region dominates in central collisions [9-12], and the transverse energy from
this region is strongly affected by the collective motion [13]. There has
never been any unambiguous experimental evidence supporting that $\sqrt{E_{t}}$
is proportional to the temperature and in fact the evidence is to the contrary.
Recent temperature measurements using both the excited states populations
and isotope yield ratios show that the temperature dependence on the impact
parameter determined from charge particle multiplicities is very
weak [14-16], less than 1 MeV from peripheral to central collisions. Similar
trends have also been observed for the $Ar+Au$ reactions at $E/A=35
MeV$ [17]. Since $E_{t}$ is strongly impact parameter dependent [18], these
temperature measurements thus imply that $E_{t}$ is independent of temperature
and the assumption that $T\propto$ $\sqrt{E_{t}}$ is not valid. Based on this
argument alone, the $\ln (1/p_{b})$ vs $ 1/\sqrt{E_{t}}$ plot is not the
Arrhenius plot analogous to that observed in chemical reactions. The breakdown
of the $T\propto \sqrt{E_{t}}$ assumption suggests that previous
interpretation of thermal scaling on emission probabilities, charge
distributions and azimuthal correlations [2-4, 20-23] should be re-examined.

Next, we will examine the assumptions used to extract the fragment emission
probability. There is no apriori reason for the emitted fragments to prefer
binomial statistics or Poissonian statistics. In Poissonian statistics, the
probability of emitting $n$ fragments is
\begin{equation}
P_{p}(n)=\frac{\lambda ^{n}}{n!}e^{-\lambda }
\end{equation}
where $\lambda =\langle n\rangle $ is the mean. The major difference between the binomial
and Poisson distributions is the ratio of the variance to the mean, $\sigma
^{2}/\langle n \rangle $ where$\ \sigma ^{2}/\langle n \rangle =1$ for Poisson distribution and $<1$ for
binomial distribution. It has been demonstrated that constraints from
conservation laws reduce the width of the Poisson distributions to much less
than 1 [19]. For example, if charge conservation constraint is applied to a
Poissonian distribution, Eq. (7) is modified to (see Appendix)
\begin{equation}
P_{m}(n, \alpha )=\frac{\lambda ^{n}}{n!}e^{-\lambda }e^{-\alpha
(n-\lambda )^{2}}
\end{equation}
where $\alpha $ is the charge constraint factor.

For small $\alpha $, the mean fragment multiplicity for Eqs. (7) and (8) are
nearly the same; $\langle n \rangle \approx \lambda ${\bf .} Fig. 1 shows three modified
Poisson distributions of Eq. (8) (solid and open points) for $
\lambda =3$, 6 and 10 and $\alpha =0.1$. To illustrate that distributions
such as Eq. (8) whose values of $\sigma ^{2}/\langle n \rangle $ are less than  1 can
be described by binomial distributions, we used Eqs. (4) and (5) to
determine the binomial parameters $m$ and $p_{b}$ whose values are listed in
Fig. 1. The solid and dashed lines are binomial distributions of Eq. (3).
The agreement between the two distributions is very good. However, in
this context, $m$ and $p_{b}$ are mainly fit parameters used to describe the
modified Poisson distributions of Eq. (8) and $p_{b}$ is not an
elementary emission probability. If the small values of
$\sigma ^{2}/\langle n \rangle $ reflect the constraints
of conservation laws of Eq. (8) observed in Ref. [19], the reducibility
of fragments emission to binomial distributions shown by Refs. [2-4] does
not imply any fundamental significance for the parameters $m$ and $p_{b}$
thereby extracted.

Even though the $\ln (1/p_{b})$ vs $ 1/\sqrt{E_{t}}$ plots constructed in
heavy ion reactions are not true Arrhenius plots, analyses of many systems
[2-4, 25] suggest that the approximate linearity of such plots may be universal.
To explain this appealing systematics, we
examine the correlations between the observable $E_{t}$, parameter $p_{b}$,
and the fragment multiplicity $n$ [25]. Since $E_{t}$ is obtained from the
measured energies of both the light particles and fragments, Eq. (2), the
energy and multiplicities are related by
\begin{equation}
E_t \approx (N_C-\langle n \rangle ) E^{LP}_t + \langle n \rangle E^{IMF}_t,
\end{equation}
where $N_{C}$ is the total charge particle multiplicities, $E_{t}^{LP}$ and $
E_{t}^{IMF}$ are the average transverse energy of a light particle and an
IMF, respectively. Experimentally, the dependence of
$N_{C}$ on $\langle n \rangle $ can be approximated by [9-12]
\begin{equation}
N_{C}=a'+b'\langle n \rangle
\end{equation}
where $a'$ corresponds to the typical number of light charge particles
emitted before any IMF is emitted and $b'$ is the number of light
charge particle emitted for each IMF emitted. There is
some non-linear dependence of $\langle n \rangle $ at very
high $N_{C}$, where $\langle  n \rangle $ saturates for central collision,
and at very low $N_{C}$, where fluctuations in small value of $n$ prevent
a sharp cutoff in $N_{c}$. Except for very large and small values of
$N_{c}$, Eq. (9) can be then rewritten into:
\begin{equation}
E_t = a + b\langle  n \rangle $ $= b (a/b + \langle  n \rangle )
\end{equation}
where $a=a'E_{t}^{LP} $ is the threshold transverse energy associated
with light particles emitted before any IMF and $b=(b'
-1)E_{t}^{LP} +E_{t}^{IMF} $.

The binomial fit parameter $m$ is nearly constant as a function of the
transverse energy $E_{t}$ [2-4]. Thus the plots of $\ln (1/p_{b})$ vs $ 1/
\sqrt{E_{t}}$ can now be reduced to $\ln (1/\langle  n \rangle ) [26]$ vs $1/$ $\sqrt{a/b+\langle  n \rangle }
$ according to Eqs. (4) and (11). Fig. 2 shows the dependence of $1/\langle  n \rangle $ and $
1/\sqrt{a/b+\langle  n \rangle }\ $for $\langle  n \rangle $ ranging from 0.25
to 5.0, typical values
observed in multifragmentation of heavy ion reactions [2-4,9-12]. For $a/b=0$,
the curve is concave and for $a/b=1$, the curve becomes slightly convex.
In the middle region, where $a/b\approx 0.5$, the curve is nearly linear.
Thus the linearity of the Arrhenius-type plots merely reflects the
correlation of the fragment multiplicity with itself when the value of $a/b$
is about 0.5 which naturally arises from the intrinsic linear dependence of $
\langle  n \rangle $ on$\ E_{t}.$

To illustrate the self-correlation effect in nuclear Arrhenius-type plots,
the published data for the $Ar+Au$ collisions at $E/A=110$ MeV [2, 4] are
plotted as solid points in the left panel of Fig. 3. The solid line is the
self-correlation of $1/\langle  n \rangle $ and $1/\sqrt{a/b+\langle  n
\rangle }$, scaled according to Eqs. (4) and (11) using the experimental
determined values of $m=12$ [2, 4], $a/b=0.5$, $b=213$ MeV [2, 4, 24]. The
good agreement between the data and the self correlation confirms that
the linearity observed in the Arrhenius-type plot mainly comes from the
linear dependence of $E_{t}$ on $\langle  n \rangle $ with a non-zero
offset in Eq. (11).

The non-zero value of $a$ arises from Eq. (10) because $a'$ is not zero. One
would expect that a relation between $1/\langle  n \rangle $ and $1/\sqrt{N_{c}}
$ similar to those shown in Fig. 2 should be observed. The right panel of Figure 3 shows
the plot of $1/\langle  n \rangle $ vs $ 1/\sqrt{N_{c}}$ for $Ar+Au$ reaction at $E/A=110$
MeV [2,4,24]. The linearity is of the plot is comparable to most
Arrhenius-type plots [2-4, 25]. Since $\langle  n \rangle $ and$\ N_{C}$
are much less affected by the energy resolution of the
detection device, they are better observables than $p_{b}$ and $E_{t}$ used
in the Arrhenius-type plots. Figure 3 suggests that the Arrhenius-type plots
contain essentially the same information as the much simpler plots of the
IMF multiplicity $(\langle  n \rangle )$ as a function of charge particle multiplicity $
(N_{C})$ published in many studies [9-12].

In summary, recent temperature measurements show that temperature is nearly
independent of impact parameter and therefore the transverse energy $E_{t}$
is not related to temperature. If $\sqrt{E_{t}}$ is not proportional to the
temperature, $
\ln (1/p_{b})$ vs $ 1/\sqrt{E_{t}}$ are not true Arrhenius plots.
Without invoking the interpretation of fragment emission probability in
binomial distributions or the temperature dependence of $E_{t}$, the
linearity of the Arrhenius-type plots can be reproduced from the linear
dependence of $E_{t}$ on fragment multiplicity $\langle  n \rangle $. These plots carry the
same information as $N_{C}$ vs $\langle  n \rangle $ plots and do not provide additional
information concerning the thermal scaling of fragment emissions.

{\bf Appendix}

When many independent processes contribute to the
multiplicity of certain particles, each with a~small probability,
then the multiplicity is expected to obey a~Poisson
distribution.  Within the grand canonical ensemble, a~Poisson distribution
also follows.  In general, the Poisson distribution is associated with a~large
overall system compared to the subsystem studied.
However, in heavy ion reactions, the systems studied are always finite,
constrained by conservation laws such as the
overall energy, charge, and mass conservation within the participant region.

To investigate the minimal effect of the above mentioned constraints on the
multiplicity distribution of~IMF, we first consider a~situation where
probabilities of emitting various individual particles are
independent, with multiplicities being governed by a Poisson
distribution in~the absence of any~constraint.  When
multiplicities of individual IMFs are governed by a Poisson
distribution, then the overall multiplicity of IMFs is also
governed by a Poisson distribution [27]. For simplicity,
we next impose only a~single constraint, that of the charge
conservation, and examine changes in the
multiplicity distribution of~IMF.  The~constraint
modifies the Poisson distribution to
\begin{eqnarray}
P_m(n) \propto  \sum_{\lbrace n_\nu : \nu \in
{\rm IMF} \rbrace} \delta_{n, \sum_\mu n_\mu} \,
\prod_\tau \left( P_p^\tau (n_\tau) \right) \, \sum_{Z_{oth}}
 {\cal P}_{oth} (Z_{oth}) \,
\delta_{ Z , Z_{oth} + \sum_{\rho} n_{\rho} \, z_{\rho}}
\, .
\end{eqnarray}
where $P_p^\tau$ is a~Poisson distribution for fragment~$\tau$,
${\cal P}_{oth}$ is the charge distribution of all particles
other than~IMF, and $Z$ is the charge of the emitting source,
$Z = \langle  Z_{oth} \rangle
+ \sum_{\nu} \langle  n_\nu \rangle z_{\nu}$.  If the system emits
a lot more other particles than~IMF, from the central-limit theorem,
${\cal P}_{oth}$ is expected to be close to a~Gaussian function,
\begin{equation}
{\cal P}_{oth}(Z_{oth}) \propto \exp{\left(-{(Z_{oth} -
\langle  Z_{oth} \rangle )^2 \over 2 \, \sigma^2(Z_{oth})}
\right)} \, .
\end{equation}
The dispersion in Eq. (13) should be primarily associated
with light (Z=1 and Z=2) particles in the participant
region, and, possibly, to some extent with the amount of charge that
the spectator matter carries off.  From the central-limit
theorem, given Poisson distributions, the~dispersion is then
$\sigma^2(Z_{oth})\gapproxeq
\langle  n_{\rm Z=1} \rangle + 4 \langle  n_{\rm Z=2} \rangle$.
Substituting Eq. (13) into Eq. (12) we get
\begin{equation}
P_m(n)  \propto \sum_{\lbrace n_\nu : \nu \in
{\rm IMF} \rbrace} \delta_{n, \sum_\mu n_\mu} \,
\prod_\tau \left( P_p^\tau (n_\tau) \right) \,
\exp{\left(-{\left(\sum_{\nu} (n_\nu - \langle  n_\nu \rangle)
z_{\nu}\right)^2 \over 2 \, \sigma^2(Z_{oth})}
\right)} \, .
\end{equation}
Finally, to~assess the effect of the constraining factor
in Eq. (14) we approximate the charge in the exponential
by its average value, $z_\nu \approx \langle  z \rangle =
\sum_\mu \langle  n_\mu \rangle z_\mu / \langle  n \rangle $,
and obtain
\begin{equation}
P_m(n)  \propto P_p (n) \,
\exp{\left( - \alpha (n - \langle  n\rangle)^2 \right)} \, ,
\end{equation}
where $\alpha = \langle  z \rangle^2 / 2 \, \sigma^2(Z_{oth})
\lapproxeq \langle  z \rangle^2 / 2 ( \langle  n_{\rm Z=1} \rangle
+ 4 \langle  n_{\rm Z=2} \rangle )$.
Charged particle multiplicities measured in Ref. [24] suggest $\alpha
\lapproxeq 0.2$.

The main effect of the constraint
in Eq. (15) is to narrow the multiplicity distribution,
compared to the
Poisson distribution.  Depending on correlations between
charge, mass, and energy within the rest of the system,
the~other constraints may affect the multiplicity distributions
further.

This work is supported by the National Science Foundation under Grant
numbers PHY-95-28844 and PHY-9403666.

\begin{figure}
\caption{Three probability distributions for the constrained Poisson
distributions of Eq. (8), for $\alpha =0.1$, $\lambda =3$ (solid points), $
\lambda =6$ (open points) and $\lambda =10$ (solid points).
The solid and
dashed lines are fits with the binomial distributions, Eqs. (3-5). The
corresponding fitting parameters, $m$ and $p_b$ are listed in the figure.}
\end{figure}
\begin{figure}
\caption{ Dependence of $1/n$ on $ 1/ \protect\sqrt{a/b + \langle  n \rangle }$
for $a/b=$0, 0.5 and 1.0.}
\end{figure}
\begin{figure}
\caption{ Left panel: Arrhenius-type plot for the $Ar+Au$
reaction at $E/A=110$~MeV~[2, 4]. The solid line is the
self-correlation of $1/\langle  n \rangle$ as a function of
$ 1/ \protect\sqrt{0.5 + \langle  n \rangle }$ scaled according to Eqs.~(4)
and~(11) with the experimental values of $m$=12, $b$=213 MeV [2,4,24]. Right
panel: Dependence of $1/\langle  n \rangle $ on $ 1/\protect\sqrt{N_{c}}$. }
\end{figure}


\begin{references}

\bibitem{r1} S. Arrhenius, Z. Phys. Chem., {\bf 4}, 226 (1889).
\bibitem{r1} L. G. Moretto et al., Phys. Rev. Lett. {\bf 74}, 1530 (1995).
\bibitem{r1} K. Tso et al., Phys. Lett. {\bf B\ 361}, 25 (1995).
\bibitem{r1} L. G. Moretto, R. Getti, L. Phair, K. Tso and G. J. Wozniak, LBL 39388
(1996), submitted to Phys. Rep.
\bibitem{r1} A.S. Botvina, D.H.E. Gross, Phys. Lett. {\bf B}, 344, 6 (1995)

R. Donangelo and S. Souza, preprint of Universidade Federal Do Rio De
Janeiro,{\bf \ if/ufrj/96}$.$
\bibitem{r1} J. Toke, D. K. Agnihotri, B. Djerroud, W. Skulski, and W.U. Schroder,
University of Rochester preprint, submitted to PRC Rapid Communications.
\bibitem{r1} W. G. Lynch, Ann Rev of Nucl \& Part. Sci. {\bf 37}, 493 (1987)
and references therein.

L. G. Moretto and G. J. Wozniak, Ann. Rev. of Nucl \& Part Sci, {\bf 43},
379 (1993).

D. H. E. Gross, Rep. Progr. Phys. {\bf 53}, 605 (1990).

\bibitem{r1} A. Chbihi et al., Phys. Rev. {\bf C\ 43}, 652 (1991).

A. Chbihi et al., Phys. Rev. {\bf C\ 43}, 666 (1991).

R. Wada et al., Phys. Rev. {\bf C\ 39}, 497 (1989).

G. Nebbia et al., Phys. Lett. {\bf B\ 176}, 20 (1986).

M. Gonin et al., Phys. Lett {\bf B\ 217}, 406 (1989).

\bibitem{r1} D. R. Bowman et al, Phys. Rev. Lett. {\bf 70}, 3534 (1993)

G. Peaslee et al., Phys. Rev. {\bf C\ 49}, R2271 (1994).

\bibitem{r1} L. Phair et al., Phys. Lett. {\bf B\ 285}, 10 (1992).

\bibitem{r1} K. Kwiatkowski et al., Phys. Rev. Lett. {\bf 74}, 3756 (1995).

J.C. Steckmeyer et al., Proceedings of the XXXIII$^{\rm rd}$ International Winter
Meeting on Nuclear Physics, Bormio, Italy, edited by I. Iori (Universita di
Milano, Milano, Italy, 1995).

\bibitem{r1} G.J. Kunde, Phys. Rev. Lett. {\bf 77}, 2897 (1996).

Dempsey et al, Phys. Rev. {\bf C\ 54}, 1710 (1996).

J. Toke et al., Phys. Rev. Lett. {\bf 77}, 3514 (1996).

\bibitem{r1} C. Williams et al., Phys. Rev. {\bf C56}, in press.

R. deSouza, Phys. Lett. {\bf B300}, 29 (1992).

\bibitem{r1} M. J. Huang et al., Phys. Rev. Lett. {\bf 77}, 1648 (1997)

\bibitem{r1} H. Xi et al., MSU preprint, MSUCL-1055, (1997).

\bibitem{r1} M.J. Huang, PhD Thesis, Michigan State University (1997).

\bibitem{r1} F. Zhu et al., Phys. Lett B. {\bf 282}, 299 (1992).

F. Zhu et al., Phys. Rev. {\bf C\ 52}, 784 (1992).

\bibitem{r1} L. Phair et al, Nucl. Phys. {\bf A548}, 489 (1992).

\bibitem{r1} L. Phair et al., Phys. Lett. {\bf B\ 291}, 7 (1992).

\bibitem{r1} L. Phair et al., Phys. Rev. Lett. {\bf 75}, 213 (1995).

\bibitem{r1} L.G. Moretto et al., Phys. Rev. Lett. {\bf 75}, 4186 (1995).

\bibitem{r1} A. Ferrero et al., Phys. Rev. {\bf C\ 53}, R5 (1996).

\bibitem{r1} L. Phair et al., Phys. Rev. Lett. {\bf 77}, 822 (1996).

\bibitem{r1} L. Phair et al., PhD Thesis, Michigan State University (1993)

\bibitem{r1} W. Skulski et al., proceedings of 13th Winter Workshop on
Nuclear Dynamics, Marathon, Fl, Feb 2-7, (1997).

\bibitem{r1} The mean IMF multiplicity $\langle  n \rangle =mp_{b}$ is
relatively free of experimental and
energy resolution distortions. These distortion effects affect $p_{b}$ and $
m$ in the opposite way so that the effects cancel out for the product of $
mp_{b}$ [6].

\bibitem{r1} N.G. Van Kampen, Stochastic Processes in Physics and Chemistry,
North-Holland, Amsterdam, 1981.
\end{references}
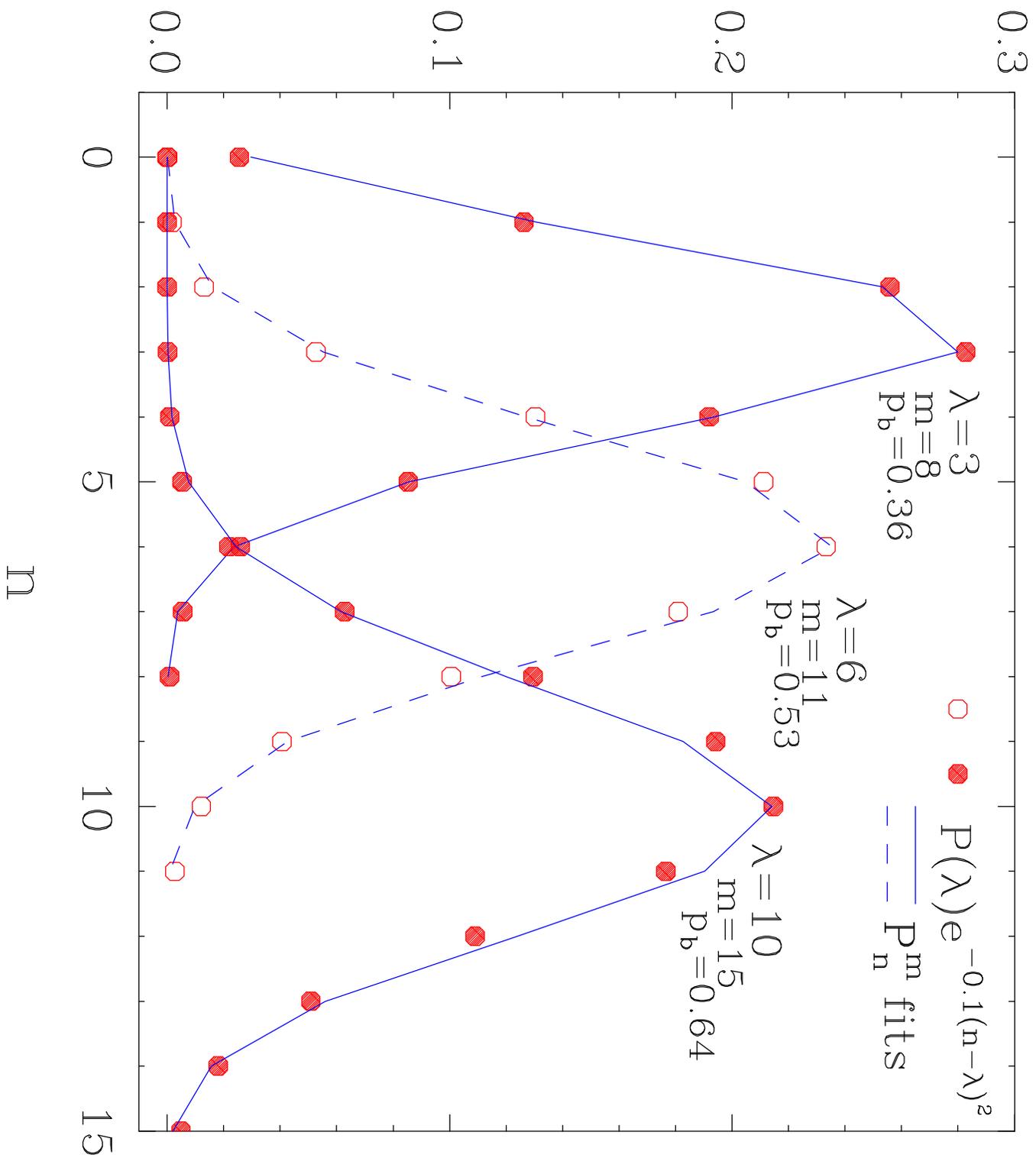
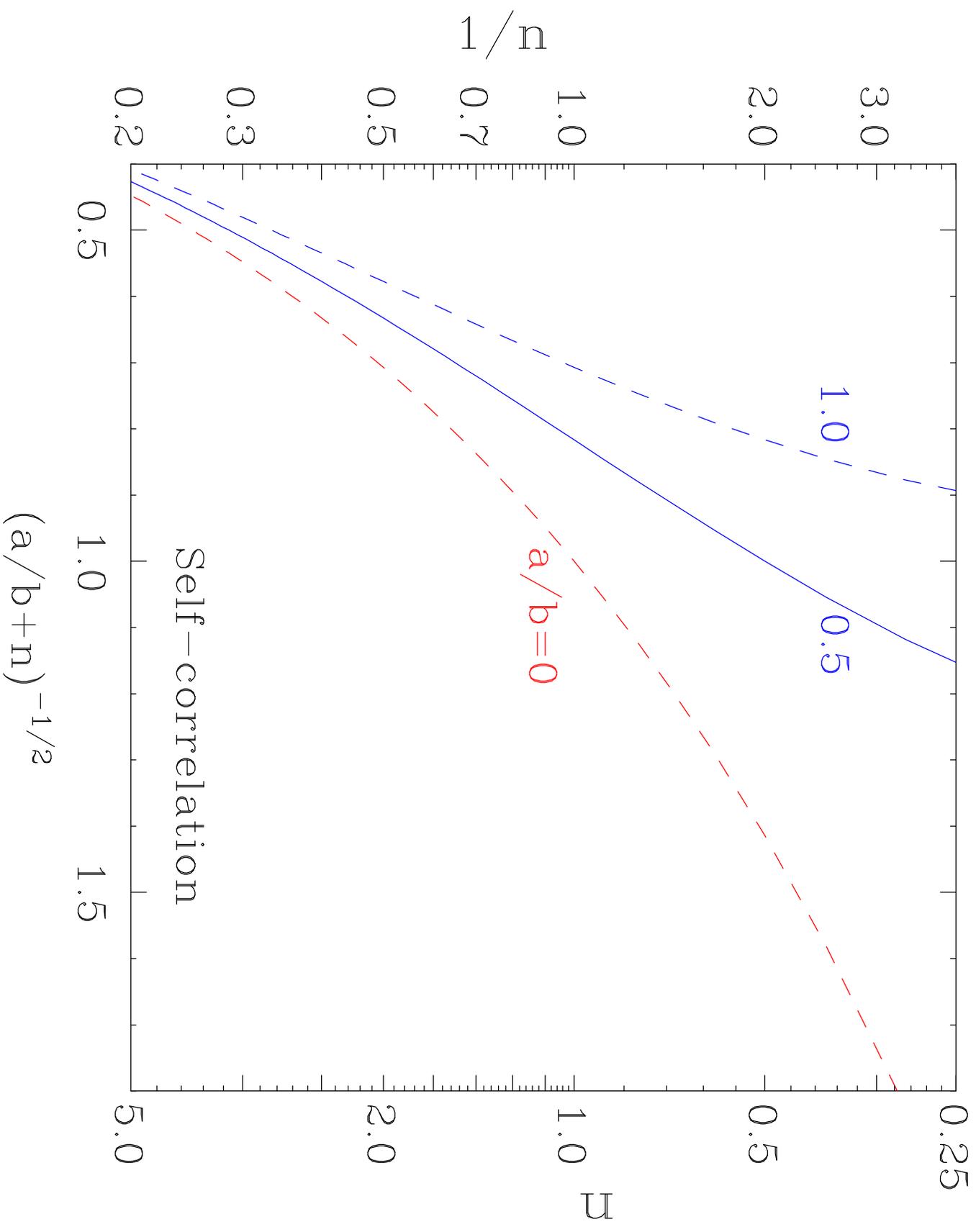
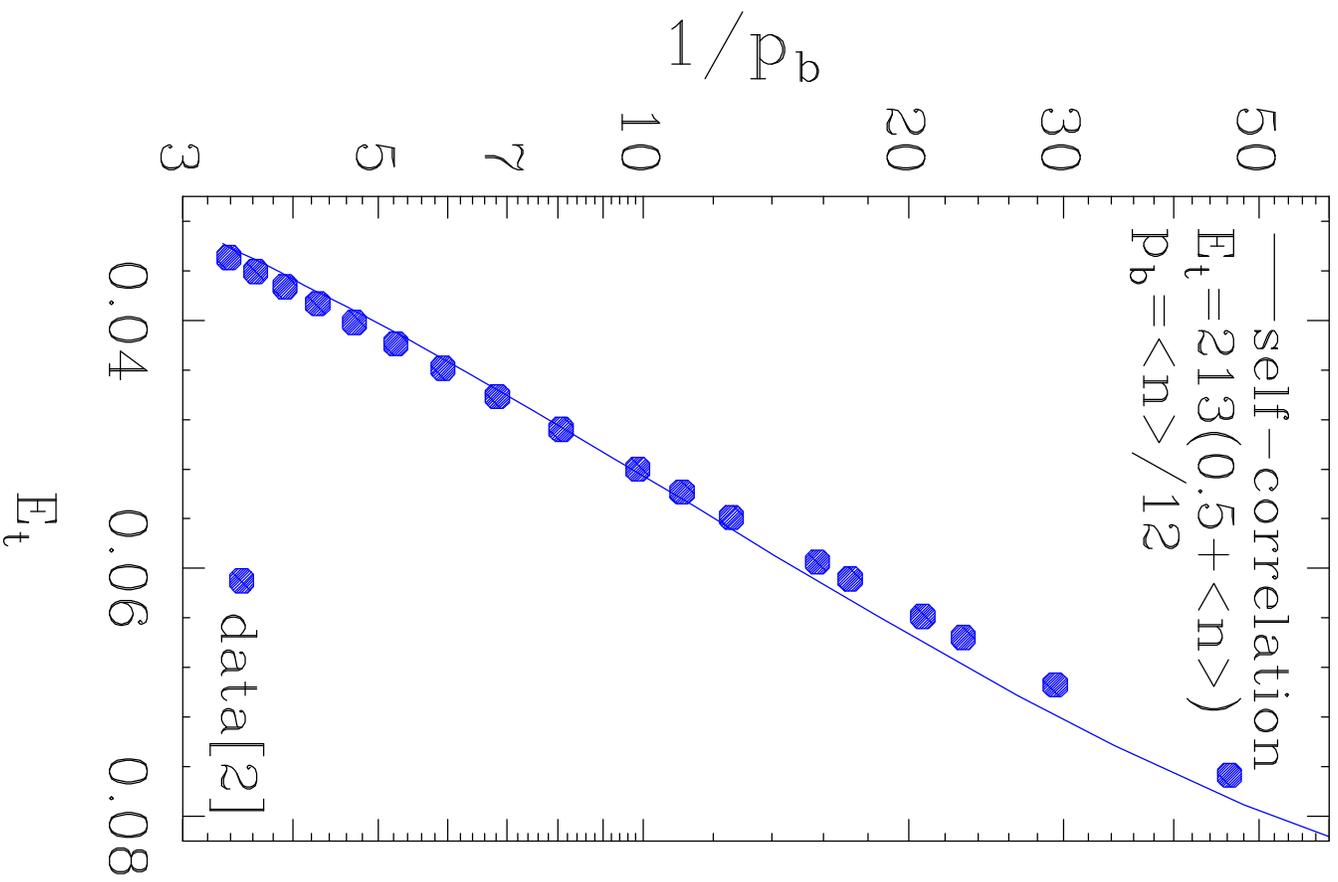
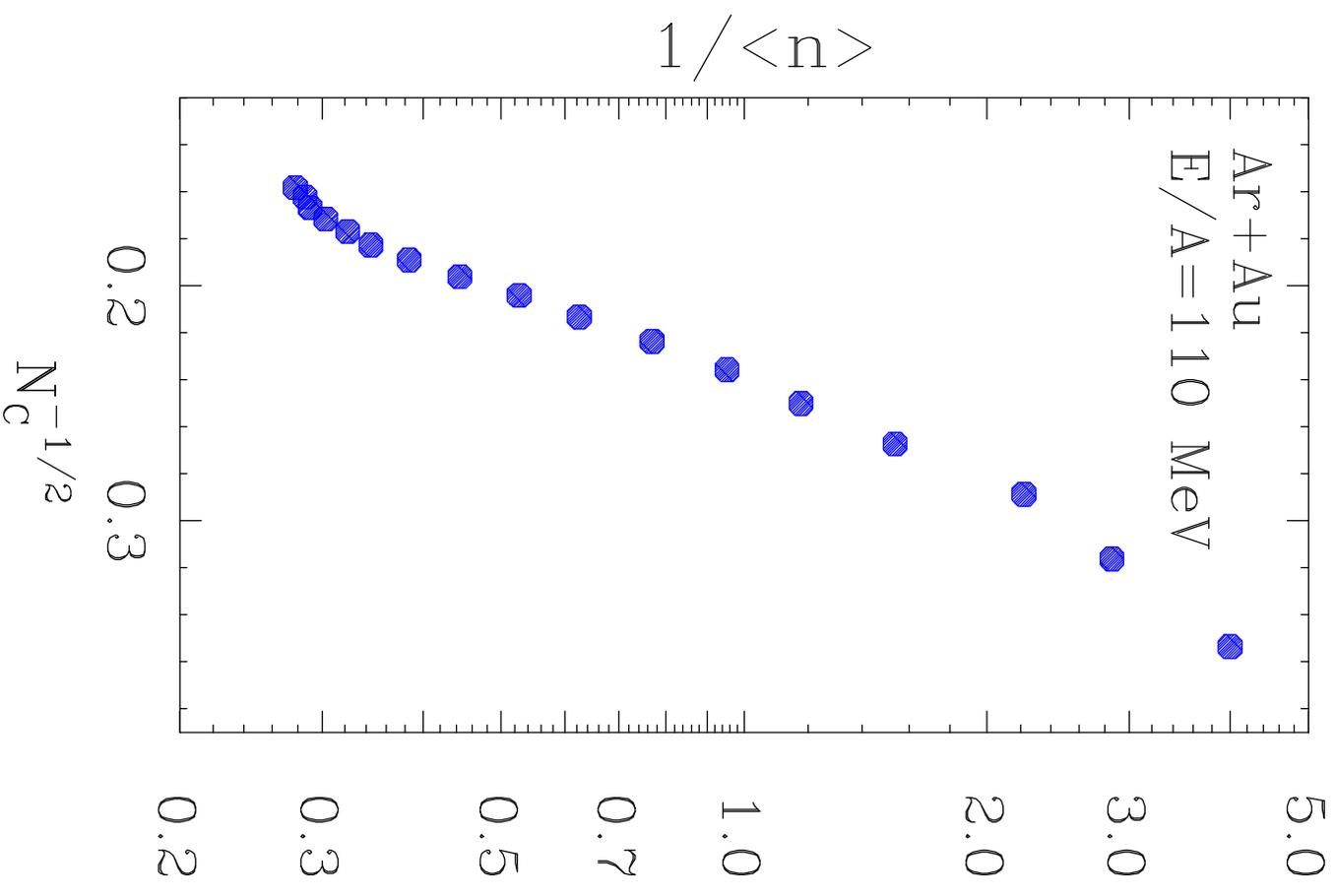
\end{document}